\documentclass[aps,preprint,nofootinbib]{revtex4-1}
\usepackage{graphicx}
\usepackage{array}
\begin{document}
\title{\Large \bf Slowly Rotating Black Hole Solutions to Ho\v{r}ava-Lifshitz Gravity}
\author{\large Alikram N. Aliev and \c{C}etin  \c{S}ent\"{u}rk}
\address{Feza G\"ursey Institute, \c Cengelk\" oy, 34684   Istanbul, Turkey}
\date{\today}

\begin{abstract}

We present a new stationary solution to the field equations of  Ho\v{r}ava-Lifshitz  gravity with the detailed balance condition and for any value of the coupling constant $\lambda > 1/3 \,$. This is the generalization of the corresponding spherically symmetric solution earlier found by L\"{u}, Mei and Pope  to include a small amount of angular momentum. For the relativistic value  $\lambda = 1 \,$, the solution describes  slowly rotating AdS type black holes.  With a soft violation of the detailed balance condition and for $\lambda = 1 $, we also find such a generalization for the Schwarzschild type black hole solution of the theory. Finally, using the canonical Hamiltonian approach, we calculate the mass and the angular momentum of these solutions.

\end{abstract}


\maketitle

\section{introduction}

General relativity  being the best classical theory of gravitation fails to describe gravity at high enough energy scales, where gravitational dynamics becomes essentially quantum in nature. This fact has led to the idea of modification of general relativity which, in physical terms, originates in the first studies of perturbative quantum gravity (see a review paper \cite{weinberg}). In the quantum context, general relativity  results in austere ultraviolet (UV) divergences in perturbation theory, defying the  renormalizability. Curing the problem of UV divergences can be achieved by adding to the Einstein-Hilbert action higher-derivative correction terms \cite{stelle}. Unfortunately,  this procedure generically entails violation of unitarity and disastrous instabilities appear in the theory \cite{woodard}.

Recently, Ho\v{r}ava proposed a new intriguing way of curing these pathologies by endowing the theory with  scaling  properties which at short distances lead to an anisotropy between space and time \cite{hora1} (see also \cite{hora2}). This type of anisotropic scalings is well known in condensed matter systems and historically traces back to the classic work of Lifshitz \cite{lif}. The degree of the anisotropy between space and time is characterized  by a number  $ z \,$, called {\it dynamical critical exponent}, the value of which is of crucial importance for the well defined UV behavior of the theory.  Ho\v{r}ava   \cite{hora1} developed a gravity model which  in the UV limit exhibits an anisotropic scaling with   $ z=3 $  fixed Lifshitz point and leads to a power-counting renormalizable theory of interacting nonrelativistic gravitons in $ 3+1 $  dimensions.  As for the relativistic invariance, it  appears as  an emergent invariance in the infrared (IR) limit through the relevant deformations of the theory, which realize the scaling with  $ z=1 $. This theory is now known as {\it Ho\v{r}ava-Lifshitz (HL) gravity} and is supposed to be a promising candidate for a UV completion of general relativity or its IR modification.

Shortly after its advent  HL  gravity has been an active arena for many investigations. Some fundamental issues related to the internal structure of the theory as well as its physically interesting extensions  were  considered in \cite{charm, svw, sibir1, rong1, tekin}.  Cosmological implications of  HL gravity have been studied in a number of papers (see for instance, \cite{kk, calc, branden, muko1}). In particular, it was shown that the higher-order spatial derivatives, which are inherent in the theory, have a drastic effect on the early history of the universe, resulting in regular cyclic and bouncing solutions \cite{kk, calc, branden}. It is also interesting that HL gravity provides a new mechanism for scale-invariant cosmological perturbations without invoking the idea of inflation \cite{kk, muko1} (see also a recent review \cite{muko2} and references therein). Another issue of fundamental significance is black hole solutions in HL gravity. The authors of \cite{lmp} found a class of static and spherically symmetric solutions with a cosmological constant. Among these solutions, the AdS type black hole solution exhibits an asymptotic behavior, which is essentially different from the Schwarzschild-AdS black hole in  general relativity. This means that general relativity  is not always recovered in  the IR limit. With a relevant deformation of the action, HL gravity  also admits an asymptotically flat static black hole solution \cite{ks}, which is a counterpart of the usual Schwarzschild black hole. Various physical properties of these  solutions  were studied \cite{cco1, cco2, lkm, park, nas}. Meanwhile, possible observational  signatures of the asymptotically flat black holes in HL gravity were examined in \cite{roman, chen1, harko} through the classical  tests of gravitational effects in both  Solar system and strong gravity regime near the black holes.

In light of these developments, it makes sense to  ask  the following  question: {\it What is the rotating counterparts of the static  black hole solutions, with or without cosmological constant, in HL gravity?} To the best of our knowledge, today this question remains largely open. There is the only case of \cite{gh}, where the authors  managed to discuss near horizon geometries of putative extremal Kerr-AdS  type black holes, for special range of parameters of the HL theory and  for special values of the parameters of these black holes. On the other hand, it is clear that in strive to answer the above question, it would be of great importance to find  genuinely HL analogs of the familiar Kerr and Kerr-AdS solutions of general relativity.
In this paper, we consider a rotating  solution to HL gravity (with the detailed balance condition) in the limit of slow rotation, but for any value of the  dynamical coupling constant $\lambda > 1/3 $. For the relativistic value  $\lambda = 1 \,$, this solution describes  slowly rotating AdS type black holes. With a soft violation of the detailed balance condition and $\lambda = 1 $, we also consider  slowly rotating asymptotically flat black holes. In Sec.II we briefly recall the basic ingredients  of HL gravity, giving the full set of the field equations, which involves an arbitrary parameter of  the soft violation. Here we also discuss  static and spherically symmetric solutions to these equations. In Sec.III we solve the field equations by linearizing them  with respect to a small rotation parameter. For this purpose, we use the metric ansatz, involving  the only  off-diagonal component governed by  a small rotation parameter.  In Sec.IV using the canonical Hamiltonian approach, we calculate the mass and the angular momentum for the slowly rotating solutions.

\section{ Ho\v{r}ava-Lifshitz Gravity and Static Black Holes}

We begin with a brief review of the basic ingredients and the field equations of HL gravity. As the basic idea  behind Ho\v{r}ava's proposal is to choose different scalings of space and time in the UV limit, it is fitting  to formulate the theory in terms of ADM type coordinates. As is known \cite{adm}, the use of such coordinates implies a (3+1) decomposition of the spacetime metric in the form
\begin{eqnarray}
ds^2 & = &-N^2 dt^2+ g_{ij}\left(dx^i+ N^i dt\right)\left(dx^j + N^j dt\right),
\label{admform}
\end{eqnarray}
where $ N $ is  the lapse function, $ N^i $ is the shift vector and  $ g_{ij} $ is the three-dimensional spatial metric. In HL gravity \cite{hora1}
two conditions were invoked: (i) the ``detailed balance" condition, which implies that the potential term in the action originates from Euclidean
three-dimensional topological massive gravity, (ii) the ``projectability" condition, requiring that the lapse function does not depend on spatial coordinates. In our description below, as in a number of cases before (see for instance, \cite{lmp,ks}) we will relax both of these conditions.

With the metric decomposition given in equation (\ref{admform}), the Einstein-Hilbert action is written in the form
\begin{equation}
I_{EH}=\frac{1}{16\pi
G}\int dt d^3x \sqrt{g}\, N \left(K_{ij}K^{ij}-K^2+R-2\Lambda\right),
\label{ehact}
\end{equation}
where $ G $ is the Newtonian gravitational constant, $ R= g^{ij}R_{ij} $ is the three-dimensional Ricci scalar, $ \Lambda $ is the cosmological constant. The extrinsic curvature of the ADM decomposition is given by
\begin{equation}
K_{ij}=\frac{1}{2N}\left(\dot{g}_{ij}-D_i N_j-D_j N_i\right),~~~~~K=g^{ij}K_{ij}\,,
\label{ext}
\end{equation}
where the dot stands for the derivative with respect to time  and the covariant derivative operator $ D $ is defined with respect to the spatial metric $ g_{ij} $.

The total action of  HL gravity is given by (see \cite{hora1,ks})
\begin{eqnarray}
I &= &\int dt d^3x \sqrt{g} N \left\{\frac{2}{\kappa^2}\left(K_{ij}K^{ij}-\lambda K^2 \right) + \frac{\kappa^2 \mu^2(\Lambda_W+\nu)}{8(1-3\lambda)}\left(R-\frac{3 \Lambda_W^2}{\Lambda_W+\nu}\right)\right. \nonumber \\[2mm]  & & \left.
+\frac{\kappa^2\mu^2(1-4\lambda)}{32(1-3\lambda)}\,R^2
-\frac{\kappa^2}{2\omega^4}\left(C_{ij}-\frac{\mu\omega^2}
{2}R_{ij}\right)\left(C^{ij}-\frac{\mu\omega^2}{2}R^{ij}\right)\right\},
\label{actHL}
\end{eqnarray}
where $ \kappa $, $ \lambda $, $ \mu $, $ \omega $ are coupling constants and $ \Lambda_W $ is a three-dimensional cosmological constant and $ C^{ij} $ is the symmetric, traceless and covariantly conserved Cotton tensor defined as
\begin{equation}
C^{ij}=\frac{\epsilon^{ikl}}{\sqrt{g}}\,D_k \left(R^j_{~l}-\frac{1}{4} \delta^j_{~l}R \right).
\label{cot}
\end{equation}
Here we have also introduced the additional  parameter
\begin{equation}\label{nu}
\nu =\frac{8\mu^2(1-3\lambda)}{\kappa^2}\,,
\end{equation}
characterizing a soft violation of the detailed balance condition \cite{ks}. This leads to the modification of the IR behavior of the theory to allow a Minkowski vacuum solution  for $ \Lambda_W \rightarrow 0 $.  Next, rescaling the time coordinate as $ t  \rightarrow c t $ and comparing the IR limit of action (\ref{actHL}) with that given in equation (\ref{ehact}), where $ \lambda =1 $, we deduce that the speed of light, the Newtonian constant and the cosmological constant appear as emergent quantities which are given by the relations
\begin{equation}
c=\frac{\kappa^2\mu}{4}\sqrt{\frac{\Lambda_W+\nu}{1-3\lambda}}\,\,,
~~~~G=\frac{\kappa^2c^2}{32\pi}\,,~~~~ \Lambda=\frac{3}{2}\,\frac{\Lambda_W^2}{\Lambda_W+ \nu}\,\,.
\label{emergent}
\end{equation}
We recall that $ \lambda $  is the dynamical coupling constant of HL gravity which governs the contribution to the theory coming from the trace of the extrinsic curvature. From equation (\ref{emergent}) it follows that the sign of  the cosmological constant is sensitive to the value of $ \lambda $. We see that  for $ \lambda > 1/3 $ it must be negative. However, for the same $ \lambda $, as noted in \cite{lmp}, one can also consider  the positive cosmological constant by making an analytical continuation with $ \mu \rightarrow i \mu ,~~ \omega^2 \rightarrow - i \omega^2 $. In what follow, we shall consider the case of the negative cosmological constant.

The field equations of HL gravity are obtained by varying action (\ref{actHL}) with respect to variables $ N \,$, $ N^i $ and $g^{ij}$. The result consists of the Hamiltonian constraint
\begin{equation}\label{ham}
-\frac{2}{\kappa^2}\left(K_{ij}K^{ij}-\lambda K^2\right)+\frac{\kappa^2\mu^2(\Lambda_W+\nu)}{8(1-3\lambda)}\left(R- \frac{3\Lambda_W^2}{\Lambda_W+\nu}\right)
+\frac{\kappa^2\mu^2(1-4\lambda)}{32(1-3\lambda)}\, R^2-\frac{\kappa^2}{2\omega^4}\,Z_{ij}Z^{ij}=0\,,
\end{equation}
and the momentum constraint
\begin{equation}\label{mom}
\frac{4}{\kappa^2}D_j \left(K^{ij}-\lambda g^{ij}K\right)=0\,,
\end{equation}
due to the variations $ \delta N $ and $ \delta N^i $, respectively  and the dynamical equation, due to the variation $ \delta g^{ij} $, which is given by
\begin{equation}
 E_{ij}=0\,,
\label{dyneq}
\end{equation}
where
\begin{equation}\label{dexp}
 E_{ij}=\frac{2}{\kappa^2}E^{(1)}_{ij}-\frac{2\lambda}{\kappa^2}E^{(2)}_{ij}
+\frac{\kappa^2\mu^2(\Lambda_W+\nu)}{8(1-3\lambda)}E^{(3)}_{ij}
+\frac{\kappa^2\mu^2(1-4\lambda)}{32(1-3\lambda)}E^{(4)}_{ij}+\frac{\kappa^2\mu}{4\omega^2}E^{(5)}_{ij}
-\frac{\kappa^2}{2\omega^4}E^{(6)}_{ij}
\end{equation}
with
\begin{eqnarray}
E^{(1)}_{ij}&=& 2 N_{(i}D_{|k|}K^k_{~j)}-2K^k_{~(i}D_{j)}N_k- N^k D_k K_{ij}
\nonumber \\[2mm]  & &
-2 N K_{ik} K^k_{~j}-\frac{1}{2}\,g_{ij} N K_{kl}K^{kl}+N K K_{ij}+\dot{K}_{ij}\,,\nonumber \\[2mm]
E^{(2)}_{ij}&=&\left(\frac{1}{2}NK^2-N^k\partial_k K+\dot{K}\right) g_{ij}+2 N_{(i}\partial_{j)}K\,, \nonumber \\[2mm]
E^{(3)}_{ij}&=&\left[R_{ij}-\frac{1}{2}g_{ij}\left(R-\frac{3\Lambda_W^2}{\Lambda_W+\nu}\right)
-D_i D_j+g_{ij} D^2
 \right]N\,,\nonumber \\[2mm]
E^{(4)}_{ij}&=& 2\left(R_{ij}-\frac{1}{4}g_{ij}R-D_i D_j+g_{ij} D^2 \right) N R \,,\nonumber \\[2mm]
E^{(5)}_{ij}&=&-2 D_k D_{(i}[Z_{j)}^{~k}N]+D^2(N Z_{ij})+g_{ij}D_k D_l(N Z^{kl})\,,\nonumber \\[2mm]
E^{(6)}_{ij}&=&\left(-\frac{1}{2}g_{ij}Z_{kl}Z^{kl}+2Z_{ik}Z^k_{~j}-2Z_{k(i}C_{j)}^{~k}+g_{ij}Z_{kl}C^{kl}\right)N
\nonumber \\[2mm] &&
-D_k[T^{kl}_{~~(i}R_{j)l}] +R^n_{~l}D_n[T^{kl}_{~~(i}g_{j)k}]\nonumber \\[2mm] &&
-D^n[T^{kl}_{~~n}g_{k(i}R_{j)l}] -D^2 D_k[T^{kl}_{~~(i}g_{j)l}]\nonumber\\[2mm] &&
+D^n[g_{l(i}D_{j)}]D_kT^{kl}_{~~n}
+D_l D_{(i}D_{|k|}T^{kl}_{~~j)}\nonumber \\[2mm] &&
+g_{ij}D^nD_k D_lT^{kl}_{~~n}\,.
\label{dyndecom}
\end{eqnarray}
Here we have used the notations $ D^2 = D_i D^i $\,,
\begin{equation}\label{z}
Z_{ij}= C_{ij}-\frac{\mu\omega^2}{2}R_{ij}\,,~~~~~T^{ij}_{~~k}=\frac{N}{\sqrt{g}}\, \epsilon^{ijl}Z_{lk}
\end{equation}
and a symmetrization procedure over the indices enclosed in round parentheses  is implied. These equations are slightly different from those obtained in \cite{lmp,kk}, but precisely  recover them for the
vanishing parameter ($\nu \rightarrow 0 $) of the soft violation of the detailed balance condition.

Exact solutions to these equations which describe static black holes  were obtained in \cite{lmp,ks}, using the standard spherically symmetric ansatz for the spacetime metric
\begin{equation}
ds^2=-\tilde{N^2}(r) f(r)\, dt^2 + \frac{dr^2}{f(r)}+r^2\left( d\theta^2+\sin^2\theta d\phi^2\right)
\label{sphermet}
\end{equation}
for which, the Cotton tensor  $ C_{ij} $ and the extrinsic curvature  $ K_{ij} $ vanish identically. In \cite{lmp}, it was shown that  with this metric ansatz the field equations given in (\ref{ham}), (\ref{mom}) and (\ref{dyneq}) admit three  different solutions. For the vanishing soft violation parameter, $\nu \rightarrow 0 $, but with arbitrary dynamical coupling $ \lambda $  one has the solution
\begin{equation}\label{sol1}
f=1-\Lambda_W r^2\,,
\end{equation}
which leaves the function $ \tilde{N} $ unconstrained. The  other two solutions are given by
\begin{equation}\label{sol23}
f=1-\Lambda_W r^2-\alpha r^{p_\pm},~~~~\tilde{N}=\beta r^{1-2{p_\pm}},
\end{equation}
where $ \alpha $ and $ \beta $ are integration constants and
 \begin{equation}\label{p}
p_\pm=\frac{2\lambda\pm\sqrt{6\lambda-2}}{\lambda-1}\,.
\end{equation}
We shall further focus on the solution, which exhibits a well-defined asymptotic behavior in the $ \lambda= 1 $ limit. This corresponds to the case with $ p = p_{-}\, $. Furthermore,  for the solution to be real, it is required that $ \lambda > 1/3 \,$. Since for $\lambda\in(1/3,\infty)$ we have $ p\in(-1,2) $\,, the $ r^2 $ term in the metric function $ f $ given in (\ref{sol23}) becomes dominant at large distances.  In the limit $ \lambda =1 $ ($p = 1/2$), we arrive at the solution  with
\begin{equation}\label{lmp1}
f =1-\Lambda_W r^2-\alpha \sqrt{r}\,,~~~~\tilde{N}=\beta\,,
\end{equation}
where the constant $\alpha $  can be related to the ADM mass of the spacetime \cite{cco2} and the constant $\beta$ can be set equal to unity by a redefinition of the time coordinate.  This gives us the static AdS type black hole solution to HL gravity \cite{lmp}. We recall that this solution was obtained within  the detailed balance condition ($\nu= 0 $). This explains the reason why the solution is significantly different from the  usual AdS-Schwarzschild  black hole solution. The authors of \cite{lmp} argued that a deviation slightly from the  detailed balance condition provides us with the AdS-Schwarzschild  black hole of HL gravity.

On the other hand, for an arbitrary soft violation parameter, $\nu \neq 0 $, there exists an asymptotically flat nonrotating black hole solution \cite{ks} for $ \lambda=1 $ and $ \Lambda_W=0 $. In this case, we have
\begin{equation}\label{ks}
f=1-\nu r^2-\sqrt{\nu^2r^4+\tilde{\alpha} r},~~~~\tilde{N}=\tilde{\beta}\,.
\end{equation}
With this function, choosing $ \tilde{\alpha}=- 4 \nu G M $ and  $ \tilde{\beta}=1 $, it is easy to see that the metric in (\ref{sphermet}) becomes the standard Schwarzschild one in the limit $ 4GM/\nu r^3 \ll1 $. We see that there also exist two event horizons located at
\begin{equation}\label{}
r_\pm=GM\left[1\pm\sqrt{1+\frac{1}{2\nu (GM)^2}}\right].
\end{equation}
With equation (\ref{nu}) in mind, it follows that the requirement of cosmic censorship  results in the bound $ \nu (GM)^2\leq 1/2 \,$. In the regime of general relativity for which $ \nu (GM)^2\ll1 $, the outer horizon approaches the usual Schwarzschild radius,  $ r_+\rightarrow 2GM $, whereas, the inner horizon shrinks to the singularity at $ r  = 0 $.

\section{Slowly Rotating Solutions}

It is clear that the static and spherically symmetric solutions, which have been  discussed above, are only special cases of more general solutions with rotational dynamics.  However, such exact solutions to HL gravity  have not been found yet. On this route, it would be certainly of great interest to find the HL counterparts of the familiar Kerr and Kerr-AdS black holes. An attempt undertaken in \cite{gh} succeeded  to present only near horizon  limits of extremal Kerr-AdS  type black holes for special range of parameters of both  HL gravity and the black holes. Here we wish to consider the rotating solutions by imposing the restriction only on their angular momentum. Namely, we restrict ourselves to  slow rotation, considering only linear order in rotation parameter perturbations around the known spherically symmetric metrics \cite{lmp,ks}. In doing this, we assume that the spectrum of small gravitational perturbations of a spherically symmetric solution in HL gravity, as that of in the case of ordinary general relativity \cite{dzn}, contains the only zero mode which descends from the presence of slow rotation. This allows us to employ  the following ansatz for the stationary  metric
\begin{equation}
\label{ansatz}
ds^2=-\tilde{N^2}(r) f(r)\, dt^2+\frac{dr^2}{f(r)}+r^2\left(d\theta^2+\sin^2\theta d\phi^2\right)+2\, a g(r)\sin^2\theta dt d\phi\,,
\end{equation}
which involves  the only  off-diagonal component with a small rotation parameter $ a $.

Next, for this metric ansatz we calculate the nonvanishing  components of the spatial tensor in (\ref{dexp}). Discarding all terms involving quadratic and higher powers in $ a $, we obtain that
\begin{eqnarray}
E_{rr}&=&\frac{\kappa^2\mu^2\sqrt{f}\,\tilde{N}}{8(3\lambda-1)r^2}\left\{(\ln \tilde{N})'\left[(\lambda-1)f'-2\lambda\frac{f-1}{r}-2(\Lambda_W+\nu)r\right]+(\lambda-1)\left(f''-2\,\frac{f-1}{r^2}\right)
\right. \nonumber \\[2mm]  & & \left.
+\frac{1}{2f}\left[(2\lambda-1)\frac{(f-1)^2}{r^2}-2\lambda\frac{f-1}{r}f'
+\frac{\lambda-1}{2}f'^2+2(\Lambda_W+\nu)(1-f-r f')-3\Lambda_W^2 r^2\right]\right\},\nonumber\\[2mm]
E_{\theta\theta}&=&\frac{\kappa^2\mu^2 r f^{3/2} \tilde{N}}{16(3\lambda-1)}
\left\{\frac{\tilde{N}''}{{\tilde N}}\left[(\lambda-1)f'-2\lambda \frac{f-1}{r}-2(\Lambda_W+\nu)r\right]
\right. \nonumber \\[2mm]  & & \left.
-\frac{(\ln \tilde{N})'}{2r^2f}\left[3r f'[-2\lambda+2r^2(\Lambda_W+\nu)+r(1-\lambda)f']
\right. \right. \nonumber \\[2mm]  & & \left.  \left.
+2f[2(1-f)+5\lambda r f'+2r^2(\Lambda_W+\nu)+ 2(1-\lambda)r^2 f'']\right]
\right. \nonumber \\[2mm]  & & \left.
-\frac{1}{r^3f}\left[-1+2\lambda+3 r^4\Lambda_W^2+(3-2\lambda)f^2+\lambda r^2 (f'^2 - f'')
\right. \right. \nonumber \\[2mm]  & & \left.  \left.
+r^4(\Lambda_W+\nu)f''
+2rf'[1+r^2(\Lambda_W+\nu)+r^2(1-\lambda)f'']
\right. \right. \nonumber \\[2mm]  & & \left.  \left.
+f[-2+2r(\lambda-2)f'+\lambda r^2 f''
+r^3(1-\lambda)f''']\right]\right\},\nonumber\\[2mm]
E_{\phi\phi}&=&\sin^2\theta E_{\theta\theta}\,.
\label{Etensorcomps}
\end{eqnarray}
Here and in the following the prime denotes differentiation with respect to the radial coordinate $ r $. We note that to the first order in rotation parameter $ a $, the above expressions do not depend on $ a $ at all. It is also not difficult to check that with the metric ansatz (\ref{ansatz}), the Hamiltonian constraint  (\ref{ham})  does not contain any term involving the rotation parameter $ a $ as well.  As a consequence, we have
\begin{eqnarray}
\frac{\kappa^2\mu^2}{8(1-3\lambda)r^2}\left\{(2\lambda-1)\frac{(f-1)^2}{r^2}
-2\lambda\frac{f-1}{r}f'+\frac{\lambda-1}{2}f'^2
\right. \nonumber \\[2mm]   \left.
+2(\Lambda_W+\nu)(1-f-r f')-3\Lambda_W^2 r^2\right\}
&=&0\,.\label{ham1}
\end{eqnarray}
In obtaining the above expressions we have used the fact that in the linear approximation in $ a $, the extrinsic curvature has the only  nonvanishing component $ K_{r\phi}=\mathcal{O}(a)$. Therefore, its trace vanishes, $ K=0 $. Furthermore, in this approximation the Cotton tensor $ C_{ij } $ vanishes as well. With all this in mind, it is straightforward to check that the metric functions given in (\ref{sol23}) for $ \lambda > 1/3 $ and  $ \nu=0   $ as well as those given in (\ref{ks}) for $\nu \neq 0 \,$,  $ \lambda=1 $ and $ \Lambda_W=0 $, satisfy the dynamical equations
\begin{eqnarray}
E_{rr}&=& 0\,,~~~~E_{\theta\theta}=0 \,,~~~~E_{\phi\phi}=0\,,
\label{Ecompseqs}
\end{eqnarray}
along with the constraint equation (\ref{ham1}). Similarly, substituting
the metric form (\ref{ansatz}) in the momentum constraint (\ref{mom}) and keeping only linear in  $ a $ terms, we arrive at the equation
\begin{eqnarray}
-\frac{2 a \sqrt{f}}{\kappa^2r^4}\left[\frac{r^4}{{\tilde N}}\left(\frac{g}{r^2}\right)'\right]' &=& 0\label{mom1}.
\end{eqnarray}
We note that this equation is valid for any value of $ \lambda $, as there is no contribution from the trace of the extrinsic curvature (K=0). Integrating (\ref{mom1}), using the explicit form of ${\tilde N}(r) $ given in  (\ref{sol23}), we find that
\begin{equation}\label{gsolp}
g=\sigma r^2+\frac{\gamma}{r^{2p}}\,,
\end{equation}
where $ \sigma $ and $ \gamma$  are  constants of integration. In the limiting case $ \lambda \rightarrow 1 $, we have
\begin{equation}\label{gsol1}
g=\sigma r^2+\frac{\gamma}{r}.
\end{equation}
In the asymptotically flat case, when $\nu \neq 0 \,, \lambda=1 $ and $ \Lambda_W=0 $, one can use $\tilde{N}= const $ to integrate equation (\ref{mom1}). As a consequence, we again arrive at equation (\ref{gsol1}).

In summary, we conclude that the metric
\begin{equation}\label{sol}
ds^2=-\tilde{N}(r)^2f(r)dt^2+\frac{dr^2}{f(r)}+r^2(d\theta^2+\sin^2\theta d\phi^2)+2N_\phi(r)dtd\phi
\end{equation}
is the solution to the field equations of HL gravity linearized in rotation parameter $ a $, for any $ \lambda > 1/3 $ and with the detailed balance condition, $ \nu=0$ . The metric functions are given by
\begin{equation}\label{metfunc2}
f=1-\Lambda_W r^2-\alpha \,r^{p},~~~~\tilde{N}=\tilde{N}(\infty)\, r^{1-2p}\,,
~~~~N_\phi=g a \sin^2\theta \,,
\end{equation}
and  we recall that
\begin{equation}
\label{pf}
p=\frac{2\lambda-\sqrt{6\lambda-2}}{\lambda-1}\,.
\end{equation}
This metric generalizes  the static and  spherically symmetric solution, earlier found in \cite{lmp} to include a small amount of angular momentum. For the relativistic value of the coupling constant, $ \lambda =1 $ ($p = 1/2$), this metric describes  a slowly rotating AdS type black hole in HL gravity. Meanwhile, with a soft violation of  the detailed balance condition, $\nu \neq 0 $, the metric in (\ref{sol}) generalizes asymptotically flat  nonrotating black hole solution of \cite{ks} with
\begin{equation}\label{ks1}
f=1-\nu r^2-\sqrt{\nu^2r^4+\tilde{\alpha} r},~~~~\tilde{N}=\tilde{N}(\infty)\,,~~~~
N_\phi=g a \sin^2\theta \,,
\end{equation}
and for $\lambda=1 $ and $ \Lambda_W=0 $. Clearly, the function $ g $ can  alternatively  be written as
\begin{equation}\label{gsolp1}
g=\frac{N^\phi(\infty)}{a} r^2+\frac{\gamma}{r^{2p}}\,.
\end{equation}
In the above expressions, for a further convenience, we have renamed the constants of integration as asymptotic values $ \tilde{N}(\infty)$ and $ N^\phi(\infty)  $. Without loss of generality, we can set $ \tilde{N}(\infty)=1 $ and  $ N^\phi(\infty)=0 $. However, in the following we need to keep them general  to calculate the physical parameters of the solutions in the framework of the  Hamiltonian approach.

\section{Mass and Angular Momentum}

In this section we calculate the mass and the angular momentum of the solution in (\ref{sol}) using the canonical Hamiltonian approach \cite{btz}. This approach was earlier employed in \cite{cco1, cco2} to obtain the mass and entropy of the static AdS type black holes in HL gravity. In  the Hamiltonian approach, the action of HL gravity (\ref{actHL}) takes the form
\begin{equation}\label{actHAM}
I=\int dt d^3x(\pi^{ij}\dot{g}_{ij}-N\mathcal{H}-N^i\mathcal{H}_i)+B\,,
\end{equation}
where
\begin{eqnarray}\label{can1}
\pi^{ij}&=&\frac{2}{\kappa^2}\sqrt{g}(K^{ij}-\lambda Kg^{ij})\,,\\[3mm]
\mathcal{H}&=&\sqrt{g}\left\{\frac{2}{\kappa^2}(K_{ij}K^{ij}-\lambda K^2)-\frac{\kappa^2\mu^2(\Lambda_W+\nu)}{8(1-3\lambda)}\left(R-\frac{3 \Lambda_W^2}{\Lambda_W+\nu}\right)
\right. \nonumber \\[2mm]  & & \left.
-\frac{\kappa^2\mu^2(1-4\lambda)}{32(1-3\lambda)}R^2
+\frac{\kappa^2}{2\omega^4}Z_{ij}Z^{ij}\right\},\\[3mm]
\mathcal{H}_i &=& -2D_j\pi_i^{~j}\,,
\end{eqnarray}
and $ B $ is a surface term, which is needed for a well defined variational principle. We recall that in our case  $ \dot{g}_{ij}=0 $ and the lapse function $ N=\tilde{N}\sqrt{f} $.  We  shall focus on solution (\ref{sol}) with the metric functions given in (\ref{metfunc2}) and (\ref{gsolp1}). With this in mind, it is straightforward to show that the action in (\ref{actHAM}) can be written in the form
\begin{equation}\label{actHAM1}
I=-2\pi(t_2-t_1)\int drd\theta ~(\tilde{N}\sqrt{f}\mathcal{H}+N^\phi\mathcal{H}_\phi)+B
\end{equation}
with
\begin{eqnarray}\label{can2}
\mathcal{H}&=& \frac{\kappa^2\mu^2}{8(3\lambda-1)}\frac{\sin\theta}{\sqrt{f}}\left [\frac{\lambda-1}{2}\,b'^2
-\frac{2\lambda}{r}\,b b'+\frac{2\lambda-1}{r^2}b^2\right],\\[2mm]
\mathcal{H}_\phi & = & \frac{2 a}{\kappa^2}\,\frac{\sin^3\theta}{\tilde{N}^2}\left[\tilde{N}(r^2g''-2g)
-\tilde{N}'(r^2g'-2rg)\right],
\end{eqnarray}
where we have introduced the new function
\begin{equation}\label{lmp3}
b=1-\Lambda_W r^2-f\,.
\end{equation}
Next, taking into account that $ N^\phi= a g/r^2 $ we perform in (\ref{actHAM1}) the integration over $ \theta $. Varying the resulting expression with respect to the functions $ b $ and $ g $\,, we find that
\begin{eqnarray}
\label{varact}
\delta I&=&-(t_2-t_1)\left\{\frac{\kappa^2\mu^2\pi}{2(3\lambda-1)}\left[(\lambda-1)\tilde{N} b'\,\delta b
-\frac{2\lambda}{r}\tilde{N}b\,\delta b\right]
\right. \nonumber \\[2mm]  & & \left.
+\frac{16\pi}{3\kappa^2}
\left[\frac{g}{\tilde{N}}\,\delta g'-\frac{g'}{\tilde{N}}\,\delta g-\frac{g}{\tilde{N}^2}\left(g'-\frac{2 }{r}\,g \right)\delta\tilde{N}\right]a^2\right\}+\delta B\,,
\end{eqnarray}
where we have omitted the terms  vanishing when the equations of motion hold. Furthermore, it is clear that the contribution from the boundary term must cancel the first two terms in this expression. In the approach under consideration, the mass and angular momentum are defined as the conjugates to the asymptotic displacements $\tilde{N}(\infty)$ and $N^\phi(\infty)$. Therefore, after some manipulations, we find that  at $ r\rightarrow\infty$
\begin{equation}\label{dB1} B=(t_2-t_1)\left\{-\tilde{N}(\infty)\left(\frac{\pi \kappa^2\mu^2}{2\sqrt{6\lambda-2}}~\alpha^2\right)
+N^\phi(\infty)\left(-\frac{32\pi(p+1)}{3\kappa^2\tilde{N}(\infty)}\,
\gamma a\right)\right\}+ B_0\,,
\end{equation}
where $ B_0 $ is arbitrary constant of integration that can be set equal to zero. In obtaining this expression we have used the formulas
\begin{eqnarray}
\left[(\lambda-1)\tilde{N} b' \,\delta b\right]_\infty
&=&\frac{\tilde{N}(\infty)}{2} \left(2\lambda-\sqrt{6\lambda-2}\right)\delta \alpha^2 \,,\\[2mm]
\left[\frac{2\lambda}{r}\tilde{N} b \,\delta b\right]_\infty &= &\lambda\tilde{N}(\infty)\,\delta\alpha^2\,,\\[2mm]
\left[\frac{g}{\tilde{N}}\,\delta g'\right]_\infty &=&-2p\, \frac{N^\phi(\infty)}{\tilde{N}(\infty)a}\,\delta\gamma\,,\\[2mm]
\left[\frac{g'}{\tilde{N}}\,\delta g\right]_\infty &= & 2\,\frac{N^\phi(\infty)}{\tilde{N}(\infty)\,a}\,\delta\gamma\,,
\end{eqnarray}
\begin{eqnarray}
\left[\frac{g}{\tilde{N}^2}\left(g'-\frac{2}{r}g\right)\delta\tilde{N}\right]_\infty &= & 0\,,
\end{eqnarray}
which are obtained by means of equations (\ref{metfunc2}) and (\ref{gsolp1}). It is now easy to identify the mass and the angular momentum from (\ref{dB1}). We have
\begin{eqnarray}
M=\frac{\pi \kappa^2\mu^2}{2\sqrt{6\lambda-2}}~\alpha^2,\label{M}\\[2mm]
J=-\frac{32\pi(p+1)}{3\kappa^2\tilde{N}(\infty)}\label{J}
\,\gamma a\,,
\end{eqnarray}
where we can now set $\tilde{N}(\infty)=1$ . It also follows that one can set $ N^\phi(\infty)=0 $, as it does not contribute to the physical parameters of the solution. For $ a=0 $,  this result agrees with that found in \cite{cco1, cco2}.

Performing similar calculations for a slowly rotating asymptotically flat black hole which is described by metric (\ref{sol}) with  the functions given in (\ref{ks1}),  we find that
\begin{eqnarray}
M &=&\frac{\pi\kappa^2\mu^2}{4}\,\tilde{\alpha}\,,~~~~
J = -\frac{16\pi}{\kappa^2}\,\gamma a\,.
\label{asflatmj}
\end{eqnarray}
We note that these expressions are in agreement with those given in (\ref{M}) and (\ref{J}) for $ \lambda =1 $ ($p = 1/2$) and $\alpha^2 \rightarrow \tilde{\alpha}$. Taking into account the emergent quantities in (\ref{emergent}) along with equation (\ref{nu}) for $ \lambda=1 $, we deduce that
\begin{eqnarray}
\tilde{\alpha} &= & -4\nu GM\,,~~~~ \gamma a =-2G J\,
\label{grreg}
\end{eqnarray}
in the regime of general relativity. We note that the value of $ \tilde{\alpha} $ in (\ref{grreg}) agrees with that given in \cite{ks}. Next, choosing in the second equation of (\ref{asflatmj}) the integration constant as
\begin{equation}\label{gammaA1}
\gamma=-\frac{\kappa^4\mu^2}{64}\,\tilde{\alpha}=-\frac{\kappa^2}{16\pi}M,
\end{equation}
it is easy to see that
\begin{equation}\label{JM}
J=M a\,,
\end{equation}
just as for a slowly rotating Kerr black hole in general relativity.

\section{Conclusion}

Ho\v{r}ava-Lifshitz  gravity is a new power-counting renormalizable theory of interacting nonrelativistic gravitons in $ 3+1 $  dimensions. Many attractive features as well as many intriguing physical and cosmological implications of this theory make sense to consider it as a possible candidate for a UV completion of general relativity. Among other implications, the black hole solutions to the theory are  certainly of great interest. In the existing literature, there are a number of static and spherically symmetric black hole solutions to HL gravity. However, rotating counterparts of these solutions, including both exact Kerr and Kerr-AdS type solutions have not found yet. In this paper, as a first step towards this goal, we have found  a new stationary  solution to HL gravity  with the detailed balance condition and for any value of the coupling constant $\lambda > 1/3 \,$.  This solution generalizes the corresponding static solution of \cite{lmp} to  include a small amount of angular momentum. For the  relativistic value of the coupling constant, $\lambda = 1 $,  the  solution corresponds to a slowly rotating AdS type black hole of HL gravity. Abandoning the detailed balance condition and going over into the value of  $\lambda = 1 $, we have also found  a slowly rotating and asymptotically flat black hole solution, thereby generalizing the Schwarzschild type solution of \cite{ks}. Using  the canonical Hamiltonian approach, we have calculated the mass and the angular momentum of these solutions.

{\it Note added.} While completing this paper, the work of \cite{lee} appeared, where for the case of $ \lambda=1  $ a slowly rotating counterpart of the static and asymptotically flat  black hole solution of \cite{ks} is also discussed.

\section{ACKNOWLEDGMENT}
\c{C}. \c{S}. wishes to thank the Scientific and Technological Research Council of Turkey (T{\"U}B\.{I}TAK) for financial support under the Programme BIDEB-2218.

\end{document}